\documentclass[acmlarge]{acmart}
\usepackage{graphicx} 
\acmDOI{10.1145/3723153}
\acmJournal{TQC}
\acmVolume{}
\acmNumber{}
\acmArticle{}
\acmYear{2025}
\acmMonth{3}
\title{Beyond NISQ: The Megaquop Machine}
\author{John Preskill}
\affiliation{\institution{Institute for Quantum Information and Matter, California Institute of Technology}\city{Pasadena, CA}\country{USA}}
\date{February 2025}



\begin{document}

\begin{abstract}
Today's Noisy Intermediate-Scale Quantum (NISQ) computers have scientific value, but quantum machines with broad practical value must be protected against noise using quantum error correction and fault-tolerant protocols. Recent studies of quantum error correction on actual hardware are opening a new era of quantum information processing. Error-corrected computers capable of performing one million quantum operations or more may be realized soon, raising a compelling question for the quantum community: What are the potential uses of these megaquop machines?

\begin{center}

\textit{Keynote address at the Q2B 2024 Conference in Silicon Valley on 11 December 2024.
} 

\end{center}

\end{abstract}

\maketitle
\section{NISQ and beyond}

Quantum technology is in the NISQ era~\cite{Preskill2018quantum}. NISQ, meaning Noisy Intermediate-Scale Quantum, is a deliberately vague term. It has no precise quantitative meaning, but is intended to convey an idea: We now have quantum machines such that brute force simulation of what the quantum machine does is well beyond the reach of our most powerful existing conventional computers~\cite{morvan2024phase}. However, these quantum machines are not error-corrected, and noise severely limits their computational power.

NISQ technology already has noteworthy scientific value. But there is no proposed application of NISQ computing with commercial value for which quantum advantage has been demonstrated when compared to the best classical hardware running the best algorithms for solving the same problems. Nor are there persuasive theoretical arguments indicating that commercially viable applications will be found that do not use quantum error-correcting codes and fault-tolerant quantum computing. 
That poses a daunting challenge for quantum science and the quantum industry.

In the future we can envision FASQ\footnote{The acronym FASQ was suggested to me by Andrew Landahl.} machines, Fault-Tolerant Application-Scale Quantum computers that can run a wide variety of useful applications, but that is still a rather distant goal. What term captures the path along the road from NISQ to FASQ? Various terms retaining the ISQ format of NISQ have been proposed~\cite{arrazola2023from,severini2023bye,bacon2024acronyms}, but I would prefer to leave ISQ behind as we move forward, so I’ll speak instead of a megaquop or gigaquop machine and so on meaning one capable of executing a million or a billion quantum operations, with the understanding that mega means not precisely a million but somewhere in the vicinity of a million.

Naively, a megaquop machine would have an error rate per logical gate of order $10^{-6}$, which we don’t expect to achieve anytime soon without using error correction and fault-tolerant operation. Or maybe the logical error rate could be somewhat larger, as we expect to be able to boost the simulable circuit volume using various error mitigation techniques in the megaquop era just as we do in the NISQ era~\cite{kim2023evidence}. Importantly, the megaquop machine would be capable of achieving some tasks beyond the reach of classical, NISQ, or analog quantum devices, for example by executing circuits with of order 100 logical qubits and circuit depth of order 10,000.

What resources are needed to operate it? That depends on many things, but a rough guess is that tens of thousands of high-quality physical qubits could suffice. When will we have it? I don’t know, but if it happens in just a few years a likely modality is Rydberg atoms in optical tweezers, assuming they continue to advance in both scale and performance.

What will we do with it? I don’t know, but as a scientist I expect we can learn valuable lessons by simulating the dynamics of many-qubit systems on megaquop machines. Will there be applications that are commercially viable as well as scientifically instructive? That I can’t promise you.

\section{The road to fault tolerance}

To proceed along the road to fault tolerance, what must we achieve? We would like to see many successive rounds of accurate error syndrome measurement such that when the syndromes are decoded the error rate per measurement cycle drops sharply as the code increases in size. Furthermore, we want to decode rapidly, as will be needed to execute universal gates on protected quantum information. Indeed, we will want the logical gates to have much higher fidelity than physical gates, and for the logical gate fidelities to improve sharply as codes increase in size. We want to do all this at an acceptable overhead cost in both the number of physical qubits and the number of physical gates. And speed matters — the time on the wall clock for executing a logical gate should be as short as possible.

A snapshot of the state of the art comes from the Google Quantum AI team~\cite{acharya2024quantumerrorcorrectionsurface}. Their recently introduced Willow superconducting processor has improved transmon lifetimes, measurement errors, and leakage correction compared to its predecessor Sycamore~\cite{google2023suppressing}. With it they can perform millions of rounds of surface-code error syndrome measurement with good stability, each round lasting about a microsecond. Most notably, they find that the logical error rate per measurement round improves by a factor of 2 (a factor they call $\Lambda$) when the code distance increases from 3 to 5 and again from 5 to 7, indicating that further improvements should be achievable by scaling the device further. They performed accurate real-time decoding for the distance 3 and 5 codes. To further explore the performance of the device they also studied the repetition code, which corrects only bit flips, out to a much larger code distance. As the hardware continues to advance we hope to see larger values of $\Lambda$ for the surface code, larger codes achieving much lower error rates, and eventually not just quantum memory but also logical two-qubit gates with much-improved fidelity compared to the fidelity of physical gates.

A nagging concern has been the potential vulnerability of superconducting quantum processors to ionizing radiation such as cosmic ray muons. In these events, errors occur in many qubits at once, too many errors for the error-correcting code to fend off. To mitigate this issue, one might want to operate a superconducting processor deep underground to suppress the muon flux, or to use less efficient codes~\cite{pattison2023hierarchicalmemoriessimulatingquantum} that protect against such error bursts.

The good news is that the Google team has demonstrated that so-called gap engineering of the qubits can reduce the frequency of such error bursts by orders of magnitude~\cite{mcewen2024resistinghighenergyimpactevents}. In their studies of the repetition code they found that, in the gap-engineered Willow processor, error bursts occurred about once per hour, as opposed to once every ten seconds in their earlier hardware.  Whether suppression of error bursts via gap engineering will suffice for running deep quantum circuits in the future is not certain, but this progress is encouraging. And by the way, the origin of the error bursts seen every hour or so is not yet clearly understood, which reminds us that not only in superconducting processors but in other modalities as well we are likely to encounter mysterious and highly deleterious rare events that will need to be understood and mitigated.

\section{Real-time decoding}

Fast real-time decoding of error syndromes is important because when performing universal error-corrected computation we must frequently measure encoded blocks and then perform subsequent operations conditioned on the measurement outcomes. If it takes too long to decode the measurement outcomes, that will slow down the logical clock speed. Decoding speed may be a more serious problem for superconducting circuits than for other hardware modalities where gates can be orders of magnitude slower.

For distance 5, Google achieves a latency, meaning the time from when data from the final round of syndrome measurement is received by the decoder until the decoder returns its result, of about 63 microseconds on average. In addition, it takes about another 10 microseconds for the data to be transmitted via Ethernet from the measurement device to the decoding workstation. That’s not bad, but considering that each round of syndrome measurement takes only a microsecond, faster would be preferable, and the decoding task becomes harder as the code grows in size.

Riverlane and Rigetti have demonstrated in small experiments that the decoding latency can be reduced by running the decoding algorithm on FPGAs rather than CPUs, and by integrating the decoder into the control stack to reduce communication time~\cite{caune2024demonstratingrealtimelowlatencyquantum}. Adopting such methods may become increasingly important as we scale further. Google DeepMind has shown that a decoder trained by reinforcement learning can achieve a lower logical error rate than a decoder constructed by humans~\cite{bausch2024learning}, but it’s unclear whether that will work at scale because the cost of training rises steeply with code distance. Also, the Harvard / QuEra team has emphasized that performing correlated decoding across multiple code blocks can reduce the depth of fault-tolerant constructions~\cite{zhou2024algorithmicfaulttolerancefast}, but this also increases the complexity of decoding, raising concern about whether such a scheme will be scalable.

\section{Trading simplicity for performance}

The Google processors use transmon qubits, as do superconducting processors from IBM and various other companies and research groups. Transmons are the simplest superconducting qubits and their quality has improved steadily; we can expect further improvement with advances in materials and fabrication. But a logical qubit with very low error rate surely will be a complicated object due to the hefty overhead cost of quantum error correction. Perhaps it is worthwhile to fashion a more complicated physical qubit if the resulting gain in performance might actually simplify the operation of a fault-tolerant quantum computer in the megaquop regime or well beyond. Several versions of this strategy are being pursued.

One approach uses cat qubits, in which the encoded 0 and 1 are coherent states of a microwave resonator, well separated in phase space, such that the noise afflicting the qubit is highly biased. Bit flips are exponentially suppressed as the mean photon number of the resonator increases, while the error rate for phase flips induced by loss from the resonator increases only linearly with the photon number. The AWS team, using their recently announced Ocelot quantum chip, built a repetition code to correct phase errors for cat qubits that are passively protected against bit flips, and showed that increasing the distance of the repetition code from 3 to 5 slightly improves the logical error rate~\cite{putterman2024hardwareefficientquantumerrorcorrection}. (See also Ref.~\cite{reglade2024quantum}.)

Another helpful insight is that error correction can be more effective if we know when and where the errors occur in a quantum circuit. We can apply this idea using a dual rail encoding of the qubits. With two microwave resonators, for example, we can encode a qubit by placing a single photon in either the first resonator (the 10) state, or the second resonator (the 01 state). The dominant error is loss of a photon, causing either the 01 or 10 state to decay to 00. One can check whether the state is 00, detecting whether the error occurred without disturbing a coherent superposition of 01 and 10. In a device built by the Yale / QCI team, loss errors are detected over 99\% of the time and all undetected errors are relatively rare~\cite{chou2024superconducting}. Similar results were reported by the AWS team, encoding a dual-rail qubit in a pair of transmons instead of resonators~\cite{levine2024demonstrating}.

Another idea is encoding a finite-dimensional quantum system in a state of a resonator that is highly squeezed in two complementary quadratures, a so-called GKP encoding. This year the Yale group used this scheme to encode 3-dimensional and 4-dimensional systems with decay rate better by a factor of 1.8 than the rate of photon loss from the resonator~\cite{brock2024quantumerrorcorrectionqudits}. (See also Ref.~\cite{lachance2024autonomous}.)

A fluxonium qubit is more complicated than a transmon in that it requires a large inductance which is achieved with an array of Josephson junctions, but it has the advantage of larger anharmonicity, which has enabled two-qubit gates with better than three 9s of fidelity, as the MIT team has shown~\cite{ding2023high}.

A theoretically compelling idea is to fashion an intrinsically robust qubit based on Majorana zero modes in superconducting nanowires. This scheme, tenaciously pursued by Microsoft, requires sophisticated fabrication and has progressed slowly so far. Their recently announced Majorana 1 processor incorporates fermion-parity readout circuitry which might pave the way for the first convincing demonstration of a topologically protected qubit~\cite{microsoft2025interferometric}.

Whether such trading of simplicity for performance in quantum devices will ultimately be advantageous for scaling to large systems is still unclear. But it’s appropriate to explore such alternatives which might pay off in the long run.

\section{Error correction with atomic qubits}

We have also seen recent progress on error correction with atomic qubits, both in ion traps and optical tweezer arrays. In these platforms qubits are movable, making it possible to apply two-qubit gates to any pair of qubits in the device. This opens the opportunity to use more efficient coding schemes, and in fact logical circuits are now being executed on these platforms. The Harvard / MIT / QuEra team sampled circuits with 48 logical qubits on a 280-qubit device~\cite{bluvstein2024logical}. Atom computing and Microsoft ran an algorithm with 28 logical qubits on a 256-qubit device~\cite{reichardt2024logicalcomputationdemonstratedneutral}. Quantinuum and Microsoft prepared entangled states of 12 logical qubits on a 56-qubit device~\cite{reichardt2024demonstrationquantumcomputationerror}.

However, so far in these devices it has not been possible to perform more than a few rounds of error syndrome measurement, and the results rely on error detection and postselection. That is, circuit runs are discarded when errors are detected, a scheme that won’t scale to large circuits. Efforts to address these drawbacks are in progress. Another concern is that the atomic movement slows the logical cycle time. If all-to-all coupling enabled by atomic movement is to be used in much deeper circuits, it will be important to speed up the movement quite a lot.

\section{Toward the megaquop machine}

How can we reach the megaquop regime? More efficient quantum codes like those recently discovered by the IBM team might help~\cite{bravyi2024high}. These require geometrically nonlocal connectivity and are therefore better suited for Rydberg optical tweezer arrays than superconducting processors, at least for now. Error mitigation strategies tailored for logical circuits 
might help by boosting the circuit volume that can be simulated beyond what one would naively expect based on the logical error rate~\cite{aharonov2024why}. Recent advances from the Google team, which reduce the overhead cost of logical gates, might also be helpful~\cite{gidney2024magicstatecultivationgrowing}.

What about applications? Impactful applications to chemistry typically require rather deep circuits so are likely to be out of reach for a while yet, but applications to materials science provide a more tempting target in the near term. Taking advantage of symmetries and various circuit optimizations like the ones Phasecraft has achieved~\cite{clinton2024towards}, we might start seeing informative results in the megaquop regime or only slightly beyond.

As a scientist, I’m intrigued by what we might conceivably learn about quantum dynamics far from equilibrium by doing simulations on megaquop machines, particularly in two spatial dimensions. But when seeking quantum advantage in that arena we should bear in mind that classical methods for such simulations are also advancing impressively. (See, for example, Ref.~\cite{begušić2024realtimeoperatorevolutiondimensions} and \cite{angrisani2024classicallyestimatingobservablesnoiseless}.)

To summarize, advances in hardware, control, algorithms, error correction, error mitigation, etc. are bringing us closer to megaquop machines, raising a compelling question for our community: What are the potential uses for these machines? Progress will require innovation at all levels of the stack.  The capabilities of early fault-tolerant quantum processors will guide application development, and our vision of potential applications will guide technological progress. Advances in both basic science and systems engineering are needed. These are still the early days of quantum computing technology, but our experience with megaquop machines will guide the way to gigaquops, teraquops, and beyond and hence to widely impactful quantum utility that benefits the world.

\begin{acks}
This article is based on a keynote address delivered at the Q2B 2024 Conference in Silicon Valley. I thank Matt Johnson for inviting me to speak at Q2B for each of the past eight years, and Travis Humble for encouraging me to publish the talk.
    I also thank Dorit Aharonov, Jason Alicea, Sergio Boixo, Earl Campbell, Roland Farrell, Ashley Montanaro, Mike Newman, Will Oliver, Chris Pattison, Rob Schoelkopf, and Qian Xu for helpful comments. I gratefully acknowledge support from the U.S. Department of Energy, Office of Science, National Quantum Information Science Research Centers, Quantum Systems Accelerator, and the National Science Foundation (PHY-2317110). The Institute for Quantum Information and Matter is an NSF Physics Frontiers Center.
\end{acks}

\bibliographystyle{ACM-Reference-Format}
  \bibliography{megaquop}

\end{document}